\documentclass[12pt]{article}
\usepackage{epstopdf,amsfonts,amsmath}
\usepackage{graphicx}
\DeclareGraphicsExtensions{.eps}

\newcommand\encadremath[1]{\vbox{\hrule\hbox{\vrule\kern8pt
\vbox{\kern8pt \hbox{$\displaystyle #1$}\kern8pt}
\kern8pt\vrule}\hrule}}
\def\enca#1{\vbox{\hrule\hbox{
\vrule\kern8pt\vbox{\kern8pt \hbox{$\displaystyle #1$}
\kern8pt} \kern8pt\vrule}\hrule}}

\newcommand\figureframex[3]{
\begin{figure}[bth]
\hrule\hbox{\vrule\kern8pt
\vbox{\kern8pt \vbox{
\begin{center}
{\mbox{\epsfxsize=#1.truecm\epsfbox{#2}}}
\end{center}
\caption{#3}
}\kern8pt}
\kern8pt\vrule}\hrule
\end{figure}
}
\newcommand\figureframey[3]{
\begin{figure}[bth]
\hrule\hbox{\vrule\kern8pt
\vbox{\kern8pt \vbox{
\begin{center}
{\mbox{\epsfysize=#1.truecm\epsfbox{#2}}}
\end{center}
\caption{#3}
}\kern8pt}
\kern8pt\vrule}\hrule
\end{figure}
}

\makeatletter
\@addtoreset{equation}{section}
\makeatother
\newtheorem{theorem}{Theorem}[section]

\newtheorem{remark}{Remark}[section]
\newtheorem{proposition}{Proposition}[section]
\newtheorem{lemma}{Lemma}[section]
\newtheorem{corollary}{Corollary}[section]
\newtheorem{definition}{Definition}[section]
\def\br{\begin{remark}\rm\small}
\def\er{\end{remark}}
\def\bt{\begin{theorem}}
\def\et{\end{theorem}}
\def\bd{\begin{definition}}
\def\ed{\end{definition}}
\def\bp{\begin{proposition}}
\def\ep{\end{proposition}}
\def\bl{\begin{lemma}}
\def\el{\end{lemma}}
\def\bc{\begin{corollary}}
\def\ec{\end{corollary}}
\def\beaq{\begin{eqnarray}}
\def\eeaq{\end{eqnarray}}
\newcommand{\proof}[1]{{\noindent \bf proof:}\par
{#1} $\square$}

\newcommand{\beq}{\begin{equation}}
\newcommand{\eeq}{\end{equation}}
\newcommand{\bea}{\begin{eqnarray}}
\newcommand{\eea}{\end{eqnarray}}

\newcommand{\Res}{\mathop{\,\rm Res\,}}
\usepackage[pdftex]{hyperref}
\hypersetup{colorlinks,urlcolor=magenta,citecolor=red,linkcolor=blue,filecolor=black}

\begin{document}

\sloppy

\pagestyle{empty}
\begin{flushright}
IPhT-T10/187 \\
CERN-PH-TH/2010-297
\end{flushright}
\addtolength{\baselineskip}{0.20\baselineskip}
\begin{center}
\vspace{26pt}
{\large \bf {The asymptotic expansion of Tracy-Widom GUE law \\ and symplectic invariants}}
\end{center}
\vspace{26pt}
\begin{center}

{\sl G.\ Borot}\hspace*{0.05cm}\footnote{\href{mailto:gaetan.borot@cea.fr}{gaetan.borot@cea.fr}},
{\sl B.\ Eynard}\hspace*{0.05cm}\footnote{\href{mailto:bertrand.eynard@cea.fr}{bertrand.eynard@cea.fr}}

\vspace{6pt}
${}^{1}$ Institut de Physique Th\'{e}orique de Saclay,\\
F-91191 Gif-sur-Yvette Cedex, France

\vspace{0.1cm}

${}^{2}$ CERN, Theory Division \\
CH-1211 Geneva 23, Switzerland

\end{center}

\vspace{20pt}
\begin{center}
{\bf Abstract}
\end{center}

%

\vspace{0.5cm}
We establish the relation between two objects: an integrable system related to Painlev\'{e} II equation, and the symplectic invariants of a certain plane curve $\Sigma_{TW}$. This curve describes the average eigenvalue density of a random hermitian matrix spectrum near a hard edge (a bound for its maximal eigenvalue). This shows that the $s \rightarrow -\infty$ asymptotic expansion of Tracy-Widow law $\mathrm{F}_{\mathrm{GUE}}(s)$, governing the distribution of the maximal eigenvalue in hermitian random matrices, is given by symplectic invariants.  \newpage


\vspace{26pt}
\pagestyle{plain}
\setcounter{page}{1}


\section*{Introduction}

The Tracy-Widom law $\mathrm{F}_{\mathrm{GUE}}(s)$ governs the distribution of the maximal eigenvalue in large hermitian matrices, sampled randomly in the GUE ensemble \cite{TW93}. In fact, this law is universal, it is also valid (under some hypothesis) for non gaussian ensembles in the unitary symmetry class \cite{DG}, for large Wigner matrices \cite{So} and for a certain class of large random complex covariance matrices \cite{So2}. Just like the Gumbel, Fr\'{e}chet and Weibull distribution are the possible universality class for the maximum of a large number of independent variables \cite{FT,Gne}, $\mathrm{F}_{\mathrm{GUE}}$ is a possible universality class for the maximum of strongly correlated variables. Thus, it is a very important law in physics and mathematics, and it has been observed in numerous phenomena and experiments \cite{Cuer1,Cuer2,ST}. Tracy and Widom have written in 2002 a concise review \cite{TW08} and provide article references for the situations where $\mathrm{F}_{\mathrm{GUE}}(s)$ appears.

The GUE ensemble \cite{Mehtabook} is defined as the space of $N \times N$ hermitian matrices $\mathcal{H}_N$ endowed with probability law:
\beq
\mathrm{d}\mu_N(M) = \frac{1}{\widetilde{Z}_N}\,\mathrm{d}M\,e^{-N\,\frac{Tr\,M^2}{2t}} \nonumber
\eeq
where $\mathrm{d}M$ is the canonical Lebesgue measure on the real vector space $\mathcal{H}_N$. The decomposition $M = U\:\mathrm{diag}(\lambda_1,\ldots,\lambda_N)\:U^{\dagger}$ induces the probability law on the eigenvalues:
\beq
\mathrm{d}\mu_{N}(\lambda) = \frac{1}{Z_{N}(\infty)}\prod_{1 \leq i < j \leq N} |\lambda_i - \lambda_j|^{2} \cdot \prod_{i = 1}^N \mathrm{d}\lambda_i\,e^{-N\,\frac{\lambda_i^2}{2t}}
\eeq
The expected maximum eigenvalue is $\lim_{N \rightarrow \infty}\mathbb{E}(\lambda_{\mathrm{max}}) = 2\sqrt{t}$, and the scale of fluctuations of the maximum around this value is $N^{-2/3}$. Here, we set $t = 1$. The following limit exists and defines the GUE Tracy-Widom law $\mathrm{F}_{\mathrm{GUE}}$:
\beq
\mathrm{F}_{\mathrm{GUE}}(s) = \lim_{N \rightarrow \infty} \mathrm{Prob}[\lambda_{\mathrm{max}} \leq 2 + s\,N^{-2/3}]
\eeq
In their original paper \cite{TW93}, Tracy and Widom were the first to characterize this law:
\begin{theorem}
\label{th1}Let $q(s)$ be the Hastings-McLeod solution of Painlev\'{e} II, uniquely defined by:
\beq
q'' = 2q^3 + sq,\qquad q(s)\mathop{\sim}_{s \rightarrow -\infty}\: \sqrt{-s/2} . \nonumber
\eeq
If we define $H(s) = \int_{s}^{\infty}q^2(\sigma)\,\mathrm{d}\sigma$,
\beq
\mathrm{F}_\mathrm{GUE}(s) = \exp\left(-\int_{s}^{\infty}H(\sigma)\,\mathrm{d}\sigma\right) .\nonumber
\eeq
\end{theorem}
$H(s)$ can be identified with a Hamiltonian for PII \cite{Oka}, and $\mathrm{F}_{\mathrm{GUE}}(s)$ with a $\tau$-function associated to this family of Hamiltonians (\cite{FW00} in the sense of Okamoto, \cite{BD01} in the sense of Jimbo-Miwa-Ueno).
From Theorem~\ref{th1}, $\mathrm{F}_{\mathrm{GUE}}(s)$ can be shown to have the following asymptotic expansion when $s \rightarrow -\infty$:
\beq
\mathrm{F}_{\mathrm{GUE}}(s) = C\,\exp\left(-\frac{|s|^{3}}{12} - \frac{\ln|s|}{8} + \sum_{g \geq 2} (-s/2)^{3(1 - g)}\,A_{g}\right).
\eeq
In some sense, $A_0 = -\frac{2}{3}$, $A_1 = -\frac{1}{8}$, and the higher order coefficients $A_g$ can be computed recursively thanks to Painlev\'{e} II equation. This method does not give the constant $C = 2^{1/24}\,e^{\zeta'(-1)}$, which was obtained later by Deift et al. by Riemann-Hilbert asymptotic analysis \cite{DIK}. This article aims to provide an alternative description of $A_g$:
\begin{proposition}
\label{Prop1}$A_g$ are the symplectic invariants $F^{g}(\Sigma_{\mathrm{PII}})$ associated to the spectral curve of equation:
\beq
(\Sigma_{\mathrm{PII}})\,:\qquad y^2 = \frac{1}{4}\,x^2(x^2 + 4) .\nonumber
\eeq
\end{proposition}
We will explain briefly the notion of spectral curve $\Sigma$ and symplectic invariants $F^{g}(\Sigma)$ (Section~\ref{sec:sym}). Those algebro-geometric objects were introduced axiomatically in \cite{EOFg}. One of their important property is their invariance under transformations of the spectral curve $(x,y) \rightarrow (x_o,y_o)$ such that $|\mathrm{d}x_o\wedge \mathrm{d}y_o| = |\mathrm{d}x\wedge \mathrm{d}y|$. So, we have equivalently:
\begin{proposition}
\label{Prop2}
$A_g$ are the symplectic invariants $F^{g}(\Sigma_{\mathrm{TW}})$ associated to the spectral curve of equation:
\beq
(\Sigma_{\mathrm{TW}})\,:\qquad y^2 = x + \frac{1}{x} - 2 \nonumber
\eeq
\end{proposition}

In the chronology, we first obtained Proposition~\ref{Prop2} heuristically by studying large deviations in an ensemble of random hermitian matrices conditioned by $\lambda_{\mathrm{max}} \leq a$ \cite{BEMN}. Then, we realized that the curve $\Sigma_{TW}$ is equivalent up to symplectic transformation to $\Sigma_{\mathrm{PII}}$, and the latter appears in relation with Painlev\'{e} II \cite{FIKN}. Now, we give a direct proof of Proposition~\ref{Prop1} relying on earlier results in integrable systems.

Meanwhile, we felt the need to prove the claim that Tau functions for integrable systems are related to symplectic invariants. As far as Tracy-Widom law is concerned, we only need this claim for $2 \times 2$ Lax pairs, and it was proved in a longer version of this article \cite{BETW}. We are currently working on the generalization to $d \times d$ Lax pairs.

\newpage

\subsection*{Outline}

We begin with an introduction to "symplectic invariants" and the formalism of the "topological recursion" (Section~\ref{sec:sym}).  We explain its relation to integrable systems in the case of $2 \times 2$ systems (Section~\ref{sec:1A}-\ref{sec:1B}, with Thm.~\ref{th3} and \ref{th4} as main tools). Next, we show how this can be applied to an $2 \times 2$ integrable system whose compatibility condition is the Painlev\'{e} II equation, and which admits $\mathrm{F}_{\mathrm{GUE}}(s)$ as Tau function (Section~\ref{sec:2}). This will prove Proposition~\ref{Prop1}.

\vspace{1cm}

\subsection*{Acknowledgments}

We would like to thank M.~Berg\`{e}re for sharing his computations on Schr\"{o}dinger equations and frequent valuable discussions, T.~Grava for discussions about PII, O.~Marchal for correspondence, A.~Its for comments, and G.~Schehr for motivating discussions. G.B. thanks the hospitality of the SISSA and of the Department of Maths and Statistics of Melbourne University, and the organizers of the MSRI semester on Random Matrix Theory. The work of B.E. is partly supported by the ANR project GranMa "Grandes Matrices Al\'{e}atoires" ANR-08-BLAN-0311-01, by the European Science Foundation through the Misgam program, by the Qu\'{e}bec government with the FQRNT. He also would like to thank the CRM (Centre de recherches math\'ematiques de Montr\'eal, QC, Canada) and the CERN for its hospitality.

\newpage

\section{Topological recursion and $2 \times 2$ integrable systems}
\label{sec:1}
\subsection{Axiomatic of the topological recursion}
\label{sec:sym}
Let us recall briefly what the topological recursion consists in (for a extensive review, see \cite{EORev}). It is an algorithm associating some numbers $F^{g}(\Sigma)$ and differential forms $\omega_n^{g}(\Sigma)(z_1,\ldots,z_n)$ to a regular spectral curve $\Sigma$. For our purposes, we call \emph{spectral curve} the data of:
\begin{itemize}
\item[$\bullet$] a plane curve $(\mathcal{C},x,y)$, in other words a Riemann surface $\mathcal{C}$, with $y$ and $\mathrm{d}x$ meromorphic functions  $\mathcal{C}\to\mathbb CP^1$ (In most of our examples, $\mathcal{C}=\mathbb CP^1$ is the Riemann sphere, and thus $x$ and $y$ are complex rational functions).
\item[$\bullet$] a maximal open domain $U \subseteq \mathcal{C}$ on which $x$ is a coordinate patch, called \emph{physical sheet}.
\item[$\bullet$] a Bergman kernel $B(z_1,z_2)$, i.e. a differential form in $z_1 \in \mathcal{C}$ and in $z_2 \in \mathcal{C}$, such that, in a local coordinate $\xi$:
\beq
B(z_1,z_2)  \mathop{=}_{z_1 \rightarrow z_2} \frac{\mathrm{d}\xi(z_1)\mathrm{d}\xi(z_2)}{(\xi(z_1) - \xi(z_2))^2} + O(1) \nonumber
\eeq
\end{itemize}
The zeroes of $\mathrm{d}x$ lying on $\partial U$ are called \emph{branchpoints} (name them $a_i$), and a spectral curve is said to be \emph{regular} when these zeroes are simple and $\mathrm{d}y(a_i) \neq 0$. In other words, when $\sqrt{x - x(a_i)}$ is a good coordinate around $a_i \in \mathcal{C}$ and $y - y(a_i) \propto \sqrt{x - x(a_i)}$. Before giving the full definitions of $\omega_n^{g}(\Sigma)$ and $F^g$, we need to introduce:
\begin{itemize}
\item[$\bullet$] The local conjugation. For $z$ in a neighborhood of $a_i$, there exists a unique $\overline{z} \neq z$ such that $x(z) = x(\overline{z})$.
\item[$\bullet$] The recursion kernel. We define, for $z_0 \in \mathcal{C}$, and $z$ in a neighborhood of $a_i$:
\beq
R(z_0,z) = \frac{-\frac{1}{2}\int_{z'=\overline{z}}^{z} B(z',z_0)}{\big(y(z) - y(\overline{z})\big)\mathrm{d}x(z)}
\eeq
$R$ is a differential form in $z_0$, and the inverse of a differential form in $z$.
\item[$\bullet$] For $z$ in a neighborhood of $a_i$, we introduce $\phi$ an (arbitrary) primitive of $y\mathrm{d}x$.
\end{itemize}
Then, we construct:
\beq
\omega_1^{0}(z) = -y(z)\mathrm{d}x(z),\qquad \omega_2^{0}(z_1,z_2)(\Sigma) = B(z_1,z_2)
\eeq
and by recursion on $|\chi| = 2g + n - 2 > 0$:
{\footnotesize
\bea
\omega_n^{g}(z_1,\underbrace{z_2,\ldots,z_{n }}_{z_J}) & = & \sum_{i} \Res_{\xi \rightarrow a_i} R(z_1,\xi)\left[\omega_{n + 1}^{g - 1}(\xi,\overline{\xi},z_J) + \sum_{K \subseteq J, \,0 \leq h \leq g}^{'} \omega_{|K| + 1}^{h}(\xi,z_K)\,\omega_{n - |K|}^{g - h}(\overline{\xi},z_{J\setminus K})\right] \nonumber \\
& &
\eea}
$\sum^{'}$ means that we exclude the terms where $\omega_n^{g}$ itself appears. Eventually, for $g \geq 2$:
\beq
F^g = \frac{1}{2 - 2g}\sum_{i} \Res_{z \rightarrow a_i} \phi(z)\,\omega_1^g(z)
\eeq
We refer to \cite{EOFg} for the construction of $F^1$ and $F^0$, which is more involved. Let us state the main properties of $F^{g}(\Sigma)$ and $\omega_n^{g}(\Sigma)$:
\begin{itemize}
\item[$\bullet$] For $2 - 2g - n < 0$, $\omega_n^{g}(\Sigma)\in T^*({\cal C})\otimes \dots \otimes T^*({\cal C})$, i.e. $\omega_n^{g}(\Sigma)(z_1,\ldots,z_n)$ is a meromorphic differential form in each $z_i \in \mathcal{C}$, symmetric in all $z_i$'s, and it has poles only at the branchpoints. There is no residue at these poles, and their maximal order is $2(3g + n - 2)$.
\item[$\bullet$] $F^{g}(\Sigma)$ is invariant under any transformation $(x,y) \rightarrow (x_o,y_o)$ such that $|\mathrm{d}x\wedge\mathrm{d}y| = |\mathrm{d}x_o\wedge\mathrm{d}y_o|$. For this reason, the $F^{g}(\Sigma)$ are called \emph{symplectic invariants}\footnote{However, $\omega_n^{g}(\Sigma)(z_1,\ldots,z_n)$ is shifted by an exact differential form in each $z_i$ under symplectic transformations}.
\item[$\bullet$] $\omega_n^{g}(\Sigma)$ have nice scaling properties when the spectral curve approaches a singular spectral curve. We will be more precise when needed.
\item[$\bullet$] $\omega_n^{g}(\Sigma)$ have nice properties under variation of the spectral curve.
\end{itemize}

\subsection{Relation to $2 \times 2$ integrable systems}
\label{sec:1A}
Let $\mathbf{L}(x)$ be a $2 \times 2$ matrix whose entries are rational in $x$. We consider a solution $\Psi(x)$ to the differential system:
\beq
\label{eq:psipsi}\frac{1}{N}\,\partial_x \Psi(x) = \mathbf{L}(x)\Psi(x),\qquad  \Psi = \left(\begin{array}{cc} \psi & \phi \\ \overline{\psi} & \overline{\phi}\end{array}\right)
\eeq
$N$ is a parameter of $\mathbf{L}$. The $1/N$ in front of the derivative is here for convenience, and can always be absorbed in a redefinition of $x$ and $\mathbf{L}$. It is not restrictive to assume $\mathrm{Tr}\,\mathbf{L} = 0$. So, $\partial_x\big(\mathrm{det}\,\Psi\big) = 0$, and we can choose the normalization:
\beq
\label{eq:norm}\mathrm{det}\,\Psi = 1 \nonumber
\eeq
To this system is associated the Christoffel-Darboux kernel:
\beq
\mathcal{K}(x_1,x_2) = \frac{\psi(x_1)\overline{\phi}(x_2) - \overline{\psi}(x_1)\phi(x_2)}{x_1 - x_2}
\eeq
In \cite{BE09} were also introduced correlators $\overline{\mathcal{W}}_n(x_1,\ldots,x_n)$ and connected correlators $\mathcal{W}_n(x_1,\ldots,x_n)$. The connected correlators are defined by:
\bea
\mathcal{W}_1(x) & = & \lim_{x' \rightarrow x}\left(\mathcal{K}(x,x') - \frac{1}{x - x'}\right) \nonumber \\
\mathcal{W}_2(x_1,x_2) & = & - \mathcal{K}(x_1,x_2)\mathcal{K}(x_2,x_1) - \frac{1}{(x_1 - x_2)^2} \nonumber \\
\label{eq:dee}\mathcal{W}_n(x_1,\ldots,x_n) & = & (-1)^{n + 1} \sum_{\sigma\,\mathrm{cycles}\,\mathrm{of}\,\mathfrak{S}_n} \prod_{i = 1}^n \mathcal{K}(x_i,x_{\sigma(i)})
\eea
and the correlators by:
\beq
\overline{\mathcal{W}}_n(x_1,\ldots,x_n) = "\mathrm{det}"\,\mathcal{K}(x_i,x_j)
\eeq
where "det" means that each occurence of $\mathcal{K}(x_i,x_i)$ in the determinant should be replaced by $\mathcal{W}_1(x_i)$, and each occurence of $\mathcal{K}(x_i,x_j)\mathcal{K}(x_j,x_i)$ by $-\mathcal{W}_2(x_i,x_j)$.
For example:
\beq
\mathcal{W}_1  = \psi\,\partial_x\overline{\phi} - \overline{\psi}\,\partial_x \phi = -\big(\partial_x\psi\,\overline{\phi} - \partial_x\overline{\psi}\,\phi\big) \nonumber
\eeq
It can be checked that the correlators are symmetric in the $x_i$'s, and that they do not have poles at coinciding points $x_i = x_j$, $i \neq j$.
The \emph{spectral curve of a first order differential system} is defined by the plane curve $\Sigma_N$ of equation:
\beq
\label{eq:sp}\mathrm{det}(y - \mathbf{L}(x)) =  0
\eeq
(since this is an algebraic equation, this defines a Riemann surface ${\mathcal C}$ and two meromorphic functions on it, related by the algebraic equation). As physical sheet, we take the open maximal domain $U \subseteq \mathcal{C}$ containing one preimage of the poles of $\mathbf{L}(x)$ (here $x = \infty$), and as Bergman kernel:
\beq
B(z_1,z_2) = -\mathrm{d}x(z_1)\mathrm{d}x(z_2)\,\mathcal{K}(x(z_1),x(z_2))\,\mathcal{K}(x(z_2),x(z_1))
\eeq
These objects can be studied with the following theorem.
\begin{theorem}
\label{th3}
Assume that:
\begin{itemize}
\item[$(i)$] $\mathbf{L}$ has a limit when $N \rightarrow \infty$.
\item[$(ii)$] The spectral curve $\Sigma_N$ of the system Eqn.~\ref{eq:psipsi} has a large $N$ limit $\Sigma_{\infty}$ which is regular, and has genus\footnote{A similar result holds when $\Sigma_{\infty}$ is not of genus $0$ under an extra hypothesis on $\mathcal{W}_1^{g}$. This will not be needed here.} $0$.
\item[$(iii)'$] $\mathcal{W}_1(x)$ admits an asymptotic expansion when $N \rightarrow \infty$ of the form : $\mathcal{W}_1 = \sum_{g \geq 0} N^{1 - 2g}\,\mathcal{W}_1^{g}$, and for $g\geq 1$, each coefficient $\mathcal{W}_1^{g}(x)$ for $g \neq 0$ may have singularities only at branchpoints of $\Sigma_{\infty}$.
\end{itemize}
Then, $\mathcal{W}_n(x_1,\ldots,x_n)$ admits an expansion of the form:
\beq
\label{eq:exp99} \mathcal{W}_n = \sum_{g \geq 0} N^{2 - 2g - n}\,\mathcal{W}_n^{g}
\eeq
The expansion coefficients of the correlators have only singularities at the branchpoints of $\Sigma_{\infty}$, and are computed by the topological recursion applied to $\Sigma_{\infty}$:
\bea
\mathcal{W}_n^{g}(x(z_1),\ldots,x(z_n))\mathrm{d}x(z_1)\cdots\mathrm{d}x(z_n) & = & \omega_n^{g}(\Sigma_{\infty})(z_1,\ldots,z_n) \nonumber \\ && - \delta_{g,0}\,\delta_{n,2}\,\frac{\mathrm{d}x(z_1)\,\mathrm{d}x(z_2)}{(x(z_1) - x(z_2))^2}
\eea
\end{theorem}
The result was first proved in \cite{BE09} with a stronger hypothesis $(iii)$, which was weakened to hypothesis $(iii)'$ in \cite{BETW}. Let us explain its origin in brief.
Following \cite{BBT}, we can always embed $\frac{1}{N}\,\partial_x \Psi = \mathbf{L}\Psi$ in a hierarchy of compatible differential systems, with a full family of times $\vec{t} = (t_j)_{j \in \mathbb{N}}$:
\beq
\label{eq:int} \frac{1}{N}\,\partial_x \Psi(x,\vec{t}) = \mathbf{L}(x,\vec{t})\Psi(x,\vec{t}),\qquad \frac{1}{N}\,\partial_{t_j} \Psi(x,\vec{t}) = \mathbf{M}_j(x,\vec{t})\Psi(x,\vec{t})
\eeq
Then, one can show that the expansion of Eqn.~\ref{eq:exp99} and the analytical structure of its coefficients for all $n \geq 1$ is implied by $(iii)$. The correlators $\mathcal{W}_n$ were defined such that the coefficients of the expansion $\mathcal{W}_n^g$ satisfy loop equations. Actually, the solution of loop equations with prescribed analytical structure is unique. This analytical structure is encoded in the large $N$ limit spectral curve $\Sigma_{\infty}$. And, the topological recursion was designed precisely to produce the unique solution to the loop equations with analytical structure given by $\Sigma_{\infty}$.

\subsection{Tau function and symplectic invariants}
\label{sec:1B}
It was claimed in \cite{BE09} and justified in \cite{BETW} that the Tau function of the integrable system Eqn.~\ref{eq:int} is given by a resummation of the symplectic invariants of $\Sigma_{\infty}$.

\begin{theorem}
\label{th4}
We make the following assumptions:
\begin{itemize}
\item[$(i$-$iii')$] The hypothesis of Thm.~\ref{th3}.
\item[$(iv)$] Eqn.~\ref{eq:int} has a solution $\Psi(x,\vec{t})$ which behaves like:
\beq
\Psi(x,\vec{t}) \sim \big(\mathbf{1} + o(1)\big)e^{N\mathbf{T}_{\alpha}(x,\vec{t})}
\eeq
in a given sector when $x \rightarrow x_{\alpha}$, a pole of $\mathbf{L}(x,\vec{t})$.
\item[$(v)$] $\mathbf{T}_{\alpha}(x,\vec{t})$ is a $2 \times 2$ diagonal matrix.
\end{itemize}
Then, the Jimbo-Miwa-Ueno Tau function, $\tau(\vec{t})$, associated to that solution admits has the same large $N$ asymptotic as:
\beq
\mathcal{F}(\vec{t}) = C\,\,\exp\left(\sum_{g \geq 0} N^{2 - 2g}\,F^{g}(\Sigma_{\infty}(\vec{t}))\right)
\eeq
up to a constant $C$ which does not depend on $\vec{t}$.
\end{theorem}

\section{Application to Painlev\'{e} II}
\label{sec:2}
\subsection{Integrable system associated to Painlev\'{e} II}

The Painlev\'{e} II equation $q''(s) = 2q^3(s) + sq(s)$ appears \cite{FN80} as the compatibility condition of the following system for $\Psi(x,s)$:
\beq
\label{eq:Q} \left\{\begin{array}{rcl} \partial_x \Psi & = & \mathbf{L}\,\Psi \\ \partial_s \Psi & = & \mathbf{M}\,\Psi \\ \Psi & = & \widetilde{\Psi}\,\exp\left[i\left(\frac{4}{3}x^3 + sx\right)\mathbf{\sigma}_3\right] \quad \mathrm{when}\; x \rightarrow +\infty\end{array}\right.
\eeq
where $\widetilde{\Psi} = \mathbf{1} + O(1/x)$ when $x \rightarrow +\infty$. The Lax pair $(\mathbf{L},\mathbf{M})$ is given by:
\bea
\mathbf{L}(x,s) & = & \left(\begin{array}{cc} -4ix^2 -i(s + 2q^2(s)) & 4x q(s) + 2i p(s) \\ 4x q(s) - 2i p(s) & 4ix^2 + i(s + 2q^2(s))\end{array}\right) \\
\mathbf{M}(x,s) & = & \left(\begin{array}{cc} - ix & q(s) \\ q(s) & ix \end{array}\right)
\eea
The necessary condition of existence of $\Psi$ is $\partial_{s} \mathbf{L} - \partial_{x} \mathbf{M} + [\mathbf{L},\mathbf{M}] = 0$. This implies that $q(s)$ is solution of Painlev\'{e} II, and $p(s) = q'(s)$. The asymptotic behavior of $\Psi$ (Eqn.~\ref{eq:Q}) determined the asymptotic behavior of $q(s)$, namely:
\beq
q(s) \mathop{\sim}_{s \rightarrow +\infty}\:\mathrm{Ai}(s) \sim \frac{\exp(-\frac{2}{3}s^{3/2})}{2\sqrt{\pi}s^{1/4}}
\eeq
Hastings and McLeod \cite{HMC} have shown that requiring this asymptotic determines a unique solution of Painlev\'{e} II. Moreover, this solution is also the unique one with left tail asymptotic:
\beq
\label{eq:asy} q(s) \mathop{\sim}_{s \rightarrow - \infty}\: \sqrt{-s/2}
\eeq
The existence of the solution $\Psi(x,s)$ was shown in \cite{FN80}. This system is an example of isomonodromy problem, itself closely related to a Riemann-Hilbert problem. Many authors have contributed to the study of these systems, and we refer to \cite{FIKN} for a review of the theory. In this section, our goal is to do some simple checks to put ourselves in the framework on Section~\ref{sec:1}.

 Let us introduce a redundant parameter $N$. We define:
\beq
\label{eq:ide}\left\{\begin{array}{c} x = N^{1/3}\,X \\ s = N^{2/3}\,S\end{array}\right.,\qquad q(s) = N^{1/3}Q(S)
\eeq
Then, Eqn.~\ref{eq:Q} is equivalent to:
\beq
\label{eq:Qm}  \left\{\begin{array}{rcl} \frac{1}{N}\,\partial_X\Psi & = & \mathbf{L}\,\Psi \\ \frac{1}{N}\,\partial_S \Psi & = & \mathbf{M}\,\Psi \\ \Psi & = & \widetilde{\Psi}\,\exp\left[iN\left(\frac{4}{3}X^3 + SX\right)\mathbf{\sigma}_3\right] \quad \mathrm{when}\; X \rightarrow +\infty\end{array}\right.
\eeq
with the Lax pair:
\bea
\mathbf{L}(X,S) & = & \left(\begin{array}{cc} -4iX^2 - i(S + 2Q^2(S)) & 4XQ(S) + \frac{2iQ'(S)}{N} \\ 4X Q(S) - \frac{2iQ'(S)}{N} & 4iX^2 +i(S + 2Q^2(S))\end{array}\right) \nonumber \\
\mathbf{M}(X,S) & = & \left(\begin{array}{cc} -iX & Q(S) \\ Q(S) & iX \end{array} \right)
\eea
and the compatibility equation:
\beq
\label{eq:eqQ} \frac{1}{N^2}Q''(S) = 2Q(S)^3 + SQ(S)
\eeq
Again, $Q(S)$ is given by the Hastings-McLeod solution of Painlev\'{e} II, which is such that (see Eqn.~\ref{eq:asy}) $\lim_{N \rightarrow \infty} Q(S)^2 = - \frac{S}{2}$ for any $S < 0$. The reason for introducing $N$ is that the $s \rightarrow -\infty$ asymptotics for the system Eqn.~\ref{eq:Q} corresponds to the $N \rightarrow \infty$ asymptotics for the system of Eqn.~\ref{eq:Qm}. We have now the notations of Thm.~\ref{th3}.

\subsection{Spectral curve}

After Eqn.~\ref{eq:sp}, the finite $N$ spectral curve for this system is:
\beq
Y^2 = -16X^4 - 8X^2S - S^2 - 4Q^4(S) - 4SQ^2(S) + \frac{4\big(Q'(S)\big)^2}{N^2}
\eeq
In the large $N$ limit, since $Q^2(S) \sim -S/2$ (we assume $S < 0$), we obtain:
\beq
\label{eq:sp2}(\Sigma_{\infty})\,:\qquad Y^2 = -16X^2\left(X^2 + \frac{S}{2}\right)
\eeq
It can be brought in a canonical form by rescaling:
\beq
(\Sigma_{\mathrm{PII}})\,:\qquad y^2 = \frac{1}{4}\,x^2(x^2 + 4)
\eeq
where
\beq
x = i\,\sqrt{\frac{8}{-S}}\,X,\qquad y = \frac{i}{S}\,Y
\eeq
For this transformation, $Y\mathrm{d}X = (-S/2)^{3/2}\,y\mathrm{d}x$. Going from the topological recursion of $\Sigma_{\infty}$ to that of $\Sigma_{\mathrm{PII}}$ is only a matter of rescaling:
\bea
F^{g}(\Sigma_{\infty}) & = & (-S/2)^{3(1 - g)}\,F^{g}(\Sigma_{\mathrm{PII}}) \\
\omega_n^{(g)}(\Sigma_{\infty}) & = & (-S/2)^{3(1 - g - n/2)}\,\omega_n^{g}(\Sigma_{\mathrm{PII}})
\eea
Thus:
\bea
\mathcal{F}(\Sigma_{\infty}) & = & \sum_{g \geq 0} N^{2 - 2g}\,F^{g}(\Sigma_{\infty}) \nonumber \\
& = & \sum_{g \geq 0} \big(N^{2/3}\,(-S/2)\big)^{3(1 - g)}\,F^{g}(\Sigma_{\mathrm{PII}}) \nonumber \\
\label{eq:fff}& = & \sum_{g \geq 0} (-s/2)^{3(1 - g)}\,F^{g}(\Sigma_{\mathrm{PII}})
\eea

\subsection{Proof of Proposition~\ref{Prop1}}

In our case, $\mathbf{L}(X,S)$ has only one pole, at $X = \infty$, which is already present in $\mathbf{L}^{(0)}(X,S)$. We define $Y(x)$ by Eqn.~\ref{eq:sp2} with the choice of the branch of the square root imposed by the $X \rightarrow \infty$ asymptotics in Eqn.~\ref{eq:Qm}:
\beq
Y(X) = -4iX\,\sqrt{X^2 + \frac{S}{2}},\qquad Y(x) \mathop{\sim}_{X \rightarrow +\infty} -4iX^2
\eeq
If the hypothesis of Theorem~\ref{th4} were satisfied, it would prove the identity (in the sense of $s \rightarrow -\infty$ asymptotic) between
\beq
\ln \mathrm{F}_{\mathrm{GUE}}(s) = \ln\tau(s)\qquad \mathrm{and}\qquad \mathcal{F}(\Sigma_{\infty}) = \sum_{g \geq 0} (-s/2)^{3(1 - g)}\,F^{g}(\Sigma_{\mathrm{PII}}) \nonumber
\eeq
which is the content of Proposition~\ref{Prop1}. For the moment, the state of our checklist is:
\begin{itemize}
\item[$\bullet$] Hypothesis $(i)$, $(iv)$ and $(v)$ are automatic with Eqn.~\ref{eq:Qm}.
\item[$\bullet$] $\Sigma_{\infty}$ is a regular spectral curve of genus $0$, i.e fulfils hypothesis $(ii)$. This is readily seen with the rational parametrization:
\beq
X(z) = \gamma\left(z + \frac{1}{z}\right),\qquad Y(z) = \frac{S}{2}\left(z^2 - \frac{1}{z^2}\right) \nonumber
\eeq
\item[$\bullet$] $\mathcal{W}_1$ has an expansion in odd powers of $N$ (see Lemma~\ref{Coco} below), and $\mathcal{W}_1^{g}$ has only singularities at branchpoints of $\Sigma_{\infty}$ (see Lemma~\ref{Lea} below). Hence $(iii)'$.
\end{itemize}

We now turn to the technical lemmas.

\begin{lemma}
\label{Lea}$\psi(X,S)$ (resp. $\overline{\psi}(X,S)$, $\phi(X,S)$, $\overline{\phi}(X,S)$) admits a $1/N$ expansion:
\beq
\label{eq:psi0}\psi(X,S) = \left(\widetilde{\psi}^{(0)}(X,S) +  \sum_{l \geq 1} N^{-l}\,\widetilde{\psi}^{(l)}(X,S)\right)\,\exp\left[iN\left(\frac{4}{3}X^3 + SX\right)\right] \nonumber
\eeq
and for all $l \geq 1$, $\widetilde{\psi}^{(l)}(X,S)$ has only poles at $X = \pm \sqrt{\frac{-S}{2}}$ (the branchpoints of $\Sigma_{\infty}$). In particular, it does not have singularities at $X = 0$, the other zero of $Y$.
\end{lemma}

\proof{\noindent We only sketch the proof. We already know that $Q^{(0)}(S) = \sqrt{\frac{-S}{2}}$, and the Hastings-McLeod solution of Eqn.~\ref{eq:eqQ}  has a $1/N^2$ expansion:
\beq
Q(S) = Q^{(0)}(S) + \sum_{l \geq 1} N^{-2l}\,Q^{(l)}(S) \nonumber
\eeq
Hence, $\mathbf{L}$ and $\mathbf{M}$ have a $1/N$ expansion. The large $N$ limit spectral curve $\Sigma_{\infty}$ associated to the system $\frac{1}{N}\,\partial_X \Psi = \mathbf{L}\Psi$ is:
\beq
Y(X,S) = \pm 4iX\sqrt{X^2 + S/2} \nonumber
\eeq
and $\psi$ to leading order is given by:
\bea
\psi^{(0)}(X,S) & = & \mathrm{cte}^{(0)}(S)\,\sqrt{\frac{\mathbf{L}_{12}^{(0)}(X,S)}{2Y(X)}}\,\exp\left[- N\int^X Y(\xi)\mathrm{d}\xi\right] \nonumber \\
& = & \mathrm{cte}^{(0)}(S)\,\left(\frac{-S/2}{X^2 + S/2}\right)^{1/4}\,\exp\left[\mp iN\left(\frac{4}{3}X^3 + SX\right)\right] \nonumber
\eea
To agree with the asymptotic required in Eqn.~\ref{eq:Qm}, we must choose the minus sign. On the other hand, the Iarge $N$ limit spectral curve $\underline{\Sigma}_{\infty}$ associated to the system $\frac{1}{N}\,\partial_S \Psi = \mathbf{M}\Psi$ is:
\beq
\underline{Y}(X,S) = \pm \sqrt{-\mathrm{det}\,\mathbf{M}^{(0)}} = \pm i \sqrt{X^2 + S/2} \nonumber
\eeq
Hence, $\psi$ is also given to leading order by:
\beq
\psi^{(0)}(X,S) = \underline{\mathrm{cte}}^{(0)}(X)\,\sqrt{\frac{\mathbf{M}_{12}^{(0)}(X,S)}{2\underline{Y}(X,S)}}\,\exp\left[-N\int^S \underline{Y}(X,\sigma)\mathrm{d}\sigma\right] \nonumber
\eeq
One can determine recursively the solution of each differential system (with respect to $X$, or with respect to $S$), and one finds an expansion of the form of Eqn.~\ref{eq:psi0}. By plugging this expansion alternatively in the two systems, it is possible to show that $\psi^{(l)}$ can have singularities only at common zeroes of $Y(X,S)$ and $\underline{Y}(X,S)$. In other words, the possible singularity at $X = 0$ obtained with the system with respect to $X$, is shown to disappear if we use the system with respect to $S$. At all orders in $1/N$, only the singularities at the simple branchpoints of $\Sigma_{\infty}$ remains, i.e. at $X = \pm \sqrt{-S/2}$. The discussion for $\phi$, $\overline{\phi}$ and $\overline{\psi}$ is similar.}

\begin{lemma}
\label{Coco} $\mathcal{W}_1(X)$ admits an expansion in odd powers of $1/N$:
\beq
\mathcal{W}_1(X) = \sum_{g \geq 0} N^{1 - 2g}\,\mathcal{W}_1^{g}(X) \nonumber
\eeq
and the singularities of $\mathcal{W}_1^{g}(X)$ for $g \neq 0$ are found only at branchpoints $X = \sqrt{-S/2}$.
\end{lemma}

\proof{\noindent From the asymptotics of $\psi$, we see that $\mathcal{W}_1$ has at least an expansion in powers of $1/N$, and the position of the singularities away from $X = \infty$ at all orders is a consequence of Lemma~\ref{Lea}. One can also check that a singularity at $X = \infty$ can only appear in the leading order $\mathcal{W}_1^{0}$. Let us stress the dependence in $N$ by writing $\mathbf{L}_N$, and $\Psi_N$ for the solution of Eqn.~\ref{eq:Qm}. We observe that ${}^t\mathbf{L}_{-N} = \mathbf{L}_N$, which implies that ${}^t\Psi^{-1}_{-N}$ obey the same differential system as $\Psi$. Moreover, ${}^t\Psi^{-1}_{-N}$ has the same asymptotic behavior near $x \rightarrow \infty$ as $\Psi_N$, and is also of determinant $1$. So, ${}^t\Psi_{-N}^{-1} = \Psi_N$, and at the level of the integrable kernel:
\bea
\mathcal{K}_N(x_1,x_2) & = & \frac{\psi_N(x_1)\overline{\phi}_N(x_2) - \overline{\psi}_N(x_1)\phi_N(x_2)}{x_1 - x_2} \nonumber \\
& = & \frac{\overline{\phi}_{-N}(x_1)\psi_{-N}(x_2) - \phi_{-N}(x_1)\overline{\psi}_{-N}(x_2)}{x_1 - x_2} \nonumber \\
& = & \mathcal{K}_{-N}(x_2,x_1) \nonumber
\eea
But, we can see on definition (Eqn.~\ref{eq:dee}) that the correlators take a $(-1)^n$ sign if we revert the orientation of all the cycles. Thus:
\beq
\mathcal{W}_n(x_1,\ldots,x_n)_{-N} = (-1)^n\,\mathcal{W}_n(x_1,\ldots,x_n)_{N} \nonumber
\eeq
In particular, $\mathcal{W}_1$ must be odd in $N$}

\end{document}